\documentclass[epj]{svjour}
\usepackage{graphics}
\usepackage{graphicx}
\usepackage[all]{xy}
\begin{document}
\title{Cosmological solutions from Induced Matter Model applied to 5D $f(R,T)$ gravity and the shrinking of the extra coordinate}
\author{P.H.R.S. Moraes
\thanks{\emph{e-mail:} moraes.phrs@gmail.com}%
}                     
%
%
\institute{INPE - Instituto Nacional de Pesquisas Espaciais - Divis\~ao de Astrof\'isica \\
Av. dos Astronautas 1758, S\~ao Jos\'e dos Campos, 12227-010 SP, Brazil}
\date{Received: date / Revised version: date}
%
\abstract{
In this work, I present exact cosmological solutions from Wesson's Induced Matter Model application to a general 5D metric in $f(R,T)$ theory of gravity. The non-conservation of the energy-momentum tensor, predicted by $f(R,T)$ theory, allows the derivation of a relation that describes the time evolution of the extra coordinate, revealing its compactification. It is showed that such a compactification could induce the effects of an accelerated expansion in the observable universe. 
\PACS{
     {PACS-key}{discribing text of that key}   \and
      {PACS-key}{discribing text of that key}
     } 
} 
\maketitle
\section{Introduction}
\label{sec:intro}

Nowadays the most popular cosmological model is the $\Lambda$CDM ($\Lambda$ - cold dark matter) Model, which is directly extracted from Einstein's General Relativity. It assumes, through the Friedmann-Robertson-Walker metric, the universe is homogeneous and isotropic, therefore its expansion is described by a scale factor dependent on time only. To match the cosmological observations as supernovae Ia \cite{riess/1998,perlmutter/1999} and baryon acoustic oscillations \cite{eisenstein/2005,percival/2010}, the universe in $\Lambda$CDM model needs to be filled by an exotic component, named ``dark energy" (DE), which makes its expansion to accelerate. Such an exotic component of the universe is mathematically described by the cosmological constant $\Lambda$ inserted ``by hand" in the Einstein's field equations (FEs) of General Relativity. Physically, what causes the acceleration would be the existence of a quantum vacuum energy, with negative equation of state (EoS) $p\sim-\rho$, with $p$ and $\rho$ representing the pressure and energy density of the universe, respectively. However, there is a huge discrepancy between the quantum vacuum energy values obtained from Cosmology \cite{hinshaw/2013} and from Particle Physics \cite{weinberg/1989}. Such a discrepancy raises uncertainties in $\Lambda$ physical interpretation. These uncertainties along with the coincidence problem, dark matter problem, missing satellites, hierarchy problem and other shortcomings (see \cite{clifton/2012} and references therein) yield the formulation of alternative gravitational models, from which new cosmological scenarios are obtained.

A proposal of changing standard gravity is through the consideration of extra dimensions. The Kaluza-Klein (KK) gravitational model (see \cite{overduin/1997} for a broad review) proposes the universe is empty in five dimensions (5D). It unifies gravitation and electromagnetism, through the application of the cylindrical condition - the annulment of all derivatives with respect to the extra dimension - and is considered a low-energy limit of superstring theories \cite{polchinski/1998}. 

Cosmological models derived from KK theory are continuously presented in the literature. Recently, in \cite{mukhopadhyay/2014} it was proposed a dynamic $\Lambda$ model under KK cosmology. Solutions of such an approach are in accordance with the observed features of an accelerating universe. In \cite{adhav/2011}, a KK cosmology in which polytropic gas DE is interacting with dark matter has been studied. Solutions which describe the matter-dominated universe in the far past and the DE-dominated era at late times were obtained. In \cite{purohit/2008}, a KK model is taken to study the role of the extra dimension in the expansion of the universe. Conditions for the accelerated expansion of the universe are derived.

KK theories usually admit compactified extradimensions. In fact, compactification is the only mechanism able to explain the apparent 4D nature of the universe in KK gravity. However, it is common to see such a compactification as an imposed feature of KK cosmological models (see, among many others, \cite{hinterbichler/2014,barbieri/2001,arefeva/1985}) instead of a natural characteristic of the extra coordinate evolution. I will present, in this work, a relation between the extra space-like coordinate and time, and such a relation will reveal, in a natural form, the shrinking of the extra coordinate.

Although highly uncommon, some other cosmological models have also predicted the shrinking of the extra coordinate, instead of imposing it. For instance, recently, such a feature was obtained and it was showed that it yields a free of singularity expanding universe \cite{huang/2014}.

Another trouble with compactification is that one cannot impose it arbitrarily on whichever dimensions one likes. The combination of the four non-compactified dimensions space-time plus the compactified fifth coordinate must be a solution of the higher-dimensional Einstein's FEs. Moreover, both cylindrical condition and compactification requirement are not necessarily satisfied in many KK models \cite{bejancu/2012}.

A innovative form of physically interpreting KK gravitational model, which will be applied in the present work, was brought up in \cite{wesson/1992a,wesson/1992b}, for which the properties of matter of the usual 4D universe (i.e., density and pressure) are regarded as the extra parts - due to the extra dimension - of the 5D Einstein's FEs for vacuum (remind that in KK theory, the 5D universe is empty), namely, the Wesson's Induced Matter Model (IMM). In fact, P.S. Wesson has showed that a 5D theory does not necessarily need an explicit energy-momentum tensor \cite{wesson/1988}; the extra terms of the 5D Einstein tensor may work as an induced energy-momentum tensor.

The IMM application has generated some important features on extradimensional cosmology. For instance, in \cite{mcmanus/1994} the Friedmann-Robertson-Walker cosmological models were interpreted as being purely geometrical in origin while in \cite{halpern/2002,halpern/2001} the IMM was applied to 5D anisotropic models. In \cite{israelit/2009}, the author has obtained dark matter and cosmic acceleration in 4D as induced effects of a matter free 5D bulk. Furthermore, the IMM was extended to curved spaces in \cite{romero/2013}. Such an extension has opened a number of possible applications for the theory, as a pre-big bang collapsing scenario, which was explored by the authors.

Another important reference on the subject of accelerated cosmological models obtained from 5D theories of gravity was presented in \cite{deleon/2006}, for which the author has considered the scenario where our observable universe is devised as a dynamical 4D hypersurface embedded in a 5D bulk space-time. In this model, the present cosmic acceleration is a natural consequence of such an embedding.  

Modifications on the Einstein's FEs are also presented by assuming the gravitational part of the action is given by a generic function of the Ricci scalar $R$ (remind that in standard gravity, such a function is linear in $R$), contemplating the $f(R)$ gravity theories \cite{vollick/2004,sotiriou/2010,felice/2010}. $f(R)$ static spherically symmetric solutions have been obtained in \cite{sebastiani/2011} while solutions coupled with electromagnetic field can be checked in \cite{mazharimousavi/2012}. Moreover, the authors in \cite{chakraborty/2015} have presented solutions from an extradimensional $f(R)$ model.

Recently, it was proposed a more generic gravity model, for which the action depends still on a generic function of $R$, but also on a function of $T$, the trace of the energy-momentum tensor $T_{\mu\nu}$, namely, the $f(R,T)$ theory of gravity \cite{harko/2011}. The present work will propose a cosmological model which unifies KK and $f(R,T)$ theory. Among the main features of $f(R,T)$ theory is the predicted matter-geometry coupling and the non-conservation of the energy-momentum tensor, which will be both investigated in this article.

In what concerns $f(R,T)$ cosmological models, in \cite{houndjo/2014} it was derived a Little Rip model, which reproduces the present stage of the universe dynamics and presents no singularity at future finite-time (i.e., no Big Rip). Moreover, it was shown that the second law of thermodynamics is always satisfied around such an $f(R,T)$ Little Rip universe. In \cite{sahoo/2014}, an axially symmetric space-time was considered in the presence of a perfect fluid source. The energy conditions in $f(R,T)$ gravity were studied in \cite{sharif/2013}. In \cite{shabani/2013}, the authors have obtained cosmological solutions which describe a matter-dominated scenario followed by an accelerated era. In \cite{jamil/2012}, some cosmological models were reconstructed from specific forms of $f(R,T)$ gravity. The authors in \cite{singh/2014} have proposed that the effects of a bulk viscosity in $f(R,T)$ gravity may explain the early and late time accelerations of the universe. Furthermore, it should be stressed that the $f(R,T)$ gravity authors themselves have derived, from a particular case of the $f(R,T)$ functional form, i.e., $f(R,T)=R+2f(T)$, a scale factor which predicts an accelerated expansion for the universe (see Section 3 of \cite{harko/2011}).

The points mentioned above, among others found currently in the literature, make reasonable to consider $f(R,T)$ gravity as a possible alternative to standard gravity shortcomings. Once the gravitational part of the action is generalized, including a general dependence not only on geometry but also on matter, the new terms of the derived FEs might be responsible for inducting different dynamical stages in the universe evolution, including the late-time DE era and even cosmic inflation \cite{singh/2014}. Moreover, as showed in \cite{jamil/2012}, some functional forms for $f(R,T)$ may retrieve some other cosmological models, as Chaplygin gas and quintessence, manifesting the generic aspect of such a theory of gravity, i.e., different cosmological models presented in the literature may be obtained from some particular cases of $f(R,T)$.

My proposal in this work is to extend $f(R,T)$ theory to a general 5D KK metric and obtain exact cosmological solutions from the IMM application. One might wonder the reason of applying the IMM in order to obtain the cosmological solutions. As quoted above, the $f(R,T)$ gravity predicts a coupling between geometry, through the dependence of a function of $R$ and matter, through the dependence of a function of $T$. The $T$ dependence of the gravitational lagrangian in $f(R,T)$ theory refers the geometrical origin of matter content in the universe \cite{shabani/2013}. Meanwhile, the IMM assumes the matter content of the universe is purely a geometric manifestation of a 5D empty universe. It seems reasonable, then, to apply the IMM in a 5D version of $f(R,T)$ theory. Moreover, from the non-conservation of the energy-momentum tensor, which is predicted by the theory, I will derive a relation for the evolution of the extra coordinate through time. Such a relation will induct geometrical effects in our 4D observable universe, resulting an accelerated expansion for high values of $t$. The dynamical behavior of the model is explained from some cosmological parameter calculation. 

Note that cosmological models which unify KK and $f(R,T)$ theories have already been proposed (see, for instance, \cite{reddy/2013,sahoo/2014b,moraes/2014}). However, none of them have treated general KK metrics, obtained naturally a compactified extra dimension nor investigated the non-conservation of the energy-momentum tensor, which will all be considered in this article.

The paper is organized as follows: in Section \ref{sec:frt} I present a brief review of the usual 4D $f(R,T)$ gravity and derive the FEs for such a theory, while in Section \ref{sec:5dfrt} the 5D $f(R,T)$ gravity FEs are presented. From such FEs, in Subsection \ref{ss:gs}, I derive general solutions which depend both on time and the extra coordinate while in Subsection \ref{ss:to}, solutions which depend on time only are presented. In Section \ref{sec:emtnc}, I derive, from the energy-momentum tensor non-conservation, the evolution of the extra coordinate through time. It will be showed in Subsection \ref{ss:cp}, from the calculation of some cosmological parameters for the model, that the consequences of such an evolution in the 4D observable universe are in accordance with the present accelerated expansion our universe is passing through. In Section \ref{sec:dis} I discuss the results obtained in Sections \ref{sec:5dfrt} and \ref{sec:emtnc}.

\section{$f(R,T)$ gravity}
\label{sec:frt}

In \cite{harko/2011}, Harko {\it et al.} have presented the $f(R,T)$ gravity, a theory in which the gravitational lagrangian is given by an arbitrary function of both the Ricci scalar $R$ and the trace $T$ of the energy-momentum tensor $T_{\mu\nu}$. Such a dependence on $T$ may rise from the existence of imperfect fluids or quantum effects. The variation of the model action below:

\begin{equation}\label{frt1}
S=\frac{1}{16\pi}\int f(R,T)\sqrt{-g}d^4x+\int \mathcal{L}_m\sqrt{-g}d^4x,
\end{equation}
with $f(R,T)$ representing the arbitrary function of $R$ and $T$, $g$ the determinant of the metric $g_{\mu\nu}$ with $\mu,\nu$ assuming the values $0,1,2,3$ and $\mathcal{L}_m$ the matter lagrangian density, yields the following FEs \cite{moraes/2014}:

\begin{equation}\label{frt4}
G_{\mu\nu}=8\pi T_{\mu\nu}+\lambda Tg_{\mu\nu}+2\lambda(T_{\mu\nu}+pg_{\mu\nu}).
\end{equation}
In (\ref{frt4}), $G_{\mu\nu}$ is the Einstein tensor and I have assumed $f(R,T)=R+2f(T)$ with $f(T)=\lambda T$ and $\lambda$ a constant (note also that throughout this article I will work with units such that $c=G=1$). Remind that such a functional form for $f(R,T)$ has been extensively used to obtain $f(R,T)$ solutions. For instance, in \cite{harko/2011}, it was showed that such an assumption yields a scale factor which describes an accelerated expansion for the universe while in \cite{moraes/2014} and \cite{reddy/2013b}, solutions to KK and 5D anisotropic cosmologies have been, respectively, obtained.

By applying the covariant derivative of the energy-momentum tensor in (\ref{frt4}), one obtains

\begin{equation}\label{frt5}
\nabla^{\mu}T_{\mu\nu}=\frac{2\lambda}{2\lambda-8\pi}\nabla^{\mu}(2T_{\mu\nu}+pg_{\mu\nu}),
\end{equation}
which contemplates the non-conservation of the energy-momentum tensor predicted by $f(R,T)$ theory.

Note that from Eq.(\ref{frt5}), the motion of massive test particles in $f(R,T)$ universe is non-geodesic. Moreover, due to the coupling between matter and geometry, the theory predicts an extra acceleration to be always present.

In the next section, I will construct KK $f(R,T)$ FEs from the application of the IMM in the 5D version of Eq.(\ref{frt4}). Those will be given in terms of $\rho$ and $p$, whose solutions will be presented in Subsections \ref{ss:gs} and \ref{ss:to}. 

\section{5D $f(R,T)$ field equations and their cosmological solutions from Induced Matter Model application}
\label{sec:5dfrt}

As mentioned above, the $T$-dependence of the lagrangian in $f(R,T)$ theory may be induced by exotic imperfect fluids or quantum effects. From what is reported in \cite{shabani/2013}, such an induction links with known and well accepted proposals, as geometrical curvature inducing matter and geometrical origin of the matter content in the universe (see \cite{farhoudi/2005} for a broad investigation on this topic). From this perspective, it seems reasonable and promising to interpret a 5D $f(R,T)$ gravity from the IMM \cite{wesson/1992a,wesson/1992b} point of view, since it considers the properties of matter in the 4D universe, such as density and pressure, as rising from the extradimensional geometrical terms of the vacuum FEs. 

More specifically speaking, the IMM brought up the information that an extradimensional theory does not need an explicit energy-momentum tensor \cite{wesson/1988}. In KK theory, the universe is considered empty in 5D, i.e., the FEs read

\begin{equation}\label{ex1}
G_{AB}=0,
\end{equation}
with the indices $A,B$ running from $0$ to $4$. According to IMM, by discriminating the terms on (\ref{ex1}) which depend on the extra coordinate, those can play the role of an ``induced" energy-momentum tensor. 

Let me consider a general KK metric of the form

\begin{equation}\label{5dfrt1}
ds^{2}=e^{\alpha(t,l)}dt^{2}-e^{\beta(t,l)}(dx^{2}+dy^{2}+dz^{2})-e^{\gamma(t,l)}dl^{2}.
\end{equation}
Note that in the metric above the scale factors depend not only on time, but also on the extra coordinate $l$.

By considering the metric (\ref{5dfrt1}), the application of the IMM in the 5D version of Eq.(\ref{frt4}) yields the following FEs:

\begin{equation}\label{5dfrt3}
\frac{3}{2}\left[\frac{\dot{\beta}\dot{\gamma}}{2}e^{-\alpha}+\left(\frac{\beta '\gamma '}{2}-\beta ^{'2}-\beta ''\right)e^{-\gamma}\right]=(8\pi +3\lambda)\rho -\lambda p,\
\end{equation}
\begin{equation}\label{5dfrt4}
\alpha '\dot{\beta}+\dot{\gamma}\beta '=\dot{\beta}\beta '+2\dot{\beta}',\
\end{equation}
\begin{eqnarray}\label{5dfrt5}
&&\left(\dot{\beta}\dot{\gamma}-\frac{\dot{\alpha}\dot{\gamma}}{2}+\frac{\dot{\gamma}^{2}}{2}+\ddot{\gamma}\right)e^{-\alpha}+\\\nonumber
&&\left(-\alpha '\beta '-\frac{3}{2}\beta ^{'2}+\beta '\gamma '-2\beta ''+\frac{\alpha '\gamma '}{2}-\frac{\alpha^{'2}}{2}-\alpha '' \right)e^{-\gamma}\\\nonumber
&=&2[\lambda\rho -(8\pi +3\lambda)p],\
\end{eqnarray}
\begin{equation}\label{5dfrt6}
-\frac{3}{4}(\beta^{'2}+\alpha '\beta')e^{-\gamma}=\lambda(\rho -p),
\end{equation}
in which I have considered $T_{AB}=(\rho,-p,-p,-p,0)$ and dots represent partial derivatives with respect to $t$ while primes, partial derivatives with respect to $l$. Note that differently from another IMM applications, $f(R,T)$ theory yields the 44 term on the rhs of Eq.(\ref{ex1}) to be non-null, as showed in Eq.(\ref{5dfrt6}). Although this is non-intuitive, note that despite the energy-momentum tensor has its fifth component null, the effective energy-momentum tensor on the rhs of Eq.(\ref{frt4}) still has terms proportional to $T$ and $p$.  

From \cite{wesson/1992b,halpern/2002}, the IMM application also consists in writing the time (only)-dependent FEs

\begin{equation}\label{5dfrt7}
\frac{3}{4}\dot{\beta}^{2}e^{-\alpha}+(8\pi +3\lambda)\rho -\lambda p=0,
\end{equation}
\begin{equation}\label{5dfrt8}
\left(-\frac{\dot{\alpha}\dot{\beta}}{2}+\frac{3}{4}\dot{\beta}^{2}+\ddot{\beta}\right)e^{-\alpha}-(8\pi +3\lambda)p+\lambda\rho=0,
\end{equation}
\begin{equation}\label{5dfrt9}
\frac{3}{2}\left(-\frac{\dot{\alpha}\dot{\beta}}{2}+\dot{\beta}^{2}+\ddot{\beta}\right)e^{-\alpha}+\lambda(\rho -p)=0.
\end{equation}

Therefore, from Eqs.(\ref{5dfrt3})-(\ref{5dfrt9}), note that one can derive cosmological solutions for $\rho$ and $p$ which depend on both $t$ and $l$, which I will call general solutions, and which depend on time $t$ only. Below I derive these two kinds of solutions.

\subsection{General solutions}
\label{ss:gs}

In order to solve the model FEs presented above, I will assume $\alpha=0$. Such an assumption can be interpreted merely as a rescaling of the time coordinate, without any loss of generality (see \cite{halpern/2001} for instance). In order to find a solution for $\beta$, let me use the separation of variables method, i.e., let me take $\beta=T_\beta L_\beta$ with $T_\beta$ and $L_\beta$ respectively representing functions of $t$ and $l$ only. By integrating (\ref{5dfrt4}), one obtains 

\begin{equation}\label{5dfrt10}
\gamma=\beta+2ln(T_\beta),
\end{equation}
for which I assumed the integration constant is null. Note that in \cite{middleton/2011}, the functional form of the scale factor related to the extra dimension was also found from an integration of the scale factor related to the three regular spatial dimensions. Moreover, from (\ref{5dfrt6}) and (\ref{5dfrt9}), one is able to write

\begin{equation}\label{5dfrt11}
e^{-\gamma}=2\frac{(\dot{\beta}^{2}+\ddot{\beta})}{\beta^{'2}}.
\end{equation}
Note that Eqs.(\ref{5dfrt10})-(\ref{5dfrt11}) allow us to find a differential equation for $T_\beta$ and a differential equation for $L_\beta$ if $\rho$ and $p$ are eliminated of the FEs. By plausibly assuming  $T_\beta\neq const$, those equations are

\begin{equation}\label{x1}
\ddot{T}_\beta=0,
\end{equation}
\begin{equation}\label{x2}
L_\beta ^{'2}-L_\beta L_\beta^{''}=0.
\end{equation}
Therefore, the solution for $\beta$ is

\begin{equation}\label{5dfrt12}
\beta=C_1te^{C_2l},
\end{equation}
with $C_1$ and $C_2$ being arbitrary constants. Note that solution (\ref{5dfrt12}) predicts a dependence on $ln(t)$ for $\gamma$ (Eq.(\ref{5dfrt10})). Such a behavior was also found in \cite{halpern/2001} through IMM application in anisotropic cosmological models.

By using solution (\ref{5dfrt12}), the model FEs yield, for the density of the universe,

\begin{equation}\label{5dfrt13}
\rho=\frac{3C_1e^{C_2l}}{8(4\pi+\lambda)t}(C_1te^{C_2l}-2).
\end{equation}

The substitution of Eqs.(\ref{5dfrt11}) and (\ref{5dfrt13}) in (\ref{5dfrt6}) yields, for the pressure of the universe:

\begin{equation}\label{5dfrt14}
p=\frac{3C_1e^{C_2l}}{8\lambda}\left[2C_1C_2^{2}e^{h(t,l)}+\frac{C_1te^{C_2l}-2}{(4\pi+\lambda)t}\lambda\right],
\end{equation}
with $h(t,l)\equiv -C_1te^{C_2l}+C_2l$. 

Before presenting a new kind of cosmological solutions, it might be interesting to mention that in order to solution (\ref{5dfrt13}) be physical, $C_1$ must be positive and $\lambda>-4\pi$. Such a solution will allow negative values for $C_1$ when $\lambda<-4\pi$.

\subsection{Time (only)-dependent solutions}
\label{ss:to}

From Eqs.(\ref{5dfrt7})-(\ref{5dfrt9}), one is able to write a time (only)-dependent solution for $\beta$ (and consequently for the density and for the pressure of the universe). By manipulating them, one finds

\begin{equation}\label{x3}
\left(\frac{24\pi}{\lambda}+9\right)\dot{\beta}^{2}+\left(\frac{24\pi}{\lambda}+10\right)\ddot{\beta}=0,
\end{equation}
whose solution is 

\begin{equation}\label{time1}
\beta=\left(\frac{5\pi+12}{8\pi+3\lambda}\right)\frac{2\lambda}{3\pi}ln[(24\pi^{2}+9\pi\lambda)t-\lambda C_3(24+10\lambda)],
\end{equation}
with $C_3$ being an arbitrary constant. Eq.(\ref{time1}) yields for the density of the universe

\begin{equation}\label{time2}
\rho=\frac{2\zeta(\lambda)[(3-\pi)\lambda-6\pi^{2}]}{[\pi(24\pi+9\lambda)t-2\lambda C_3(5\pi+12)]^{2}},
\end{equation}
with $\zeta(\lambda)=3(5\pi+12)\lambda/(4\pi+\lambda)$. In possession of Eqs.(\ref{time1})-(\ref{time2}), the time (only)-dependent solution for the pressure of the universe is

\begin{equation}\label{time3}
p=\frac{\zeta(\lambda)\eta(\lambda)}{[\pi(24\pi+9\lambda)t-2\lambda C_3(5\pi+12)]^{2}},
\end{equation}
with 

\begin{equation}\label{x2}
\eta(\lambda)=(\pi+24)\lambda^{3}-4\pi(5\pi-24)\lambda^{2}-2(48\pi^{3}+\pi-3)\lambda-12\pi^{2}.
\end{equation}

Note that in order to guarantee solution (\ref{time2}) to be physical, $\lambda$ must be in the range $-4\pi<\lambda<0$. Such a condition besides ensuring a physical solution for $\rho$ in (\ref{time2}), predicts an accelerated expansion for the universe, since it implies in $p<0$ in (\ref{time3}), with the sign of $C_3$ being irrelevant. 

According to what was mentioned in Section \ref{sec:intro}, in $\Lambda$CDM cosmological model, what causes the accelerated expansion of the universe is some sort of quantum vacuum energy with negative EoS, i.e., $p<0$. Such a cosmological feature is automatically obtained above just by requiring solution (\ref{time2}) to be physical.

\section{Non-conservation of the energy-momentum tensor and the time evolution of the extra coordinate}
\label{sec:emtnc}

The non-conservation of the energy-momentum tensor in $f(R,T)$ theory will have a valuable application in the 5D case, as it will be demonstrated below.

Note that by taking $A=B=0$ in the 5D version of Eq.(\ref{frt5}), one has

\begin{equation}\label{eec1}
\frac{8\pi-\lambda}{2\lambda}\dot{\rho}+\dot{p}=0,
\end{equation}
while $A=B=4$ yields

\begin{equation}\label{eec3}
\frac{p'}{p}e^{\gamma}+(e^{\gamma})'=0.
\end{equation}
The other terms will vanish since $\rho$ and $p$ do not depend on the coordinates $x,y$ and $z$.

Now let me substitute the general solutions (\ref{5dfrt13})-(\ref{5dfrt14}) in (\ref{eec1})-(\ref{eec3}). Such a procedure yields the following equations

\begin{equation}\label{emtnc1}
C_1C_2t\sqrt{e^{h(t,l)+C_2l}}=\sqrt{\frac{8\pi+\lambda}{2(4\pi+\lambda)}},
\end{equation}
\begin{equation}\label{emtnc2}
e^{-h(t,l)+C_2l}=4C_1C_2^{2}\frac{(4\pi+\lambda)}{\lambda}\frac{te^{C_2l}}{(2-C_1^{2}t^{2}e^{C_2l})}.
\end{equation}
By substituting (\ref{emtnc2}) in (\ref{emtnc1}) and solving for $l$, one obtains 

\begin{equation}\label{emtnc3}
l_i=\frac{1}{C_2}\{ln[\xi_i(t)]-ln(t)\}
\end{equation}
as solutions for $l$ as functions of $t$, with $i=1$ or $2$ and

\begin{equation}\label{emtnc4}
\xi_1=\frac{C_1\lambda-\sqrt{-16C_1^{3}\pi\lambda t+C_1^{2}\lambda^{2}-2C_1^{3}\lambda^{2} t}}{C_1^{3}\lambda^{2} t},
\end{equation}
\begin{equation}\label{emtcn5}
\xi_2=\frac{C_1\lambda+\sqrt{-16C_1^{3}\pi\lambda t+C_1^{2}\lambda^{2}-2C_1^{3}\lambda^{2}t}}{C_1^{3}\lambda^{2} t}.
\end{equation}
Figure \ref{Fig1} below shows the evolution of the extra coordinate $l$ through time from Eqs.(\ref{emtnc3})-(\ref{emtnc4}).

\begin{figure}[ht!]
\vspace{0.5cm}
\centering
\includegraphics[height=5.5cm,angle=00]{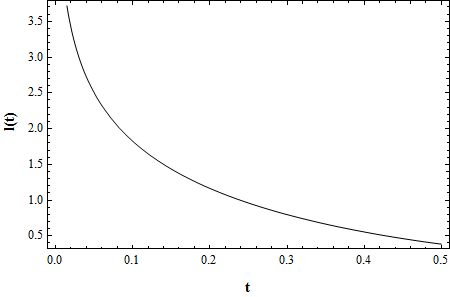}
\caption{Time evolution of $l$ from Eqs.(\ref{emtnc3})-(\ref{emtnc4}) with $C_1=-1$, $C_2=1$ and $\lambda=-5\pi$.}
\label{Fig1}
\end{figure} 
As one can see, Figure \ref{Fig1} predicts a compactification of the extra coordinate $l$ through time. This is usually assumed in KK cosmological models (see, for instance, \cite{hinterbichler/2014,barbieri/2001,arefeva/1985}), but here, the non-conservation of the energy-momentum tensor, which is an important fundamental property of $f(R,T)$ theories, has revealed such a feature, with no need of prior conjectures. 

\subsection{Cosmological parameters}
\label{ss:cp}

In this subsection, I will present the time evolution of some cosmological parameters of the present model. Those will be derived from the scale factors (\ref{5dfrt10}) and (\ref{5dfrt12}). However, for the dependence on $l$, I will use Eqs.(\ref{emtnc3})-(\ref{emtnc4}). Such a procedure allow us to verify the effects that the evolution of the extra coordinate causes in our 4D observable universe.

Note that because of the exponential in (\ref{5dfrt12}), the dependence of the cosmological parameters on $C_2$ will vanish. Therefore, I will take once again $C_1=-1$ (as in Fig.\ref{Fig1}) and write the cosmological parameters in terms of $\lambda$ and $t$ only. 

The mean scale factor $a$ tell us how the expansion of the universe depends on time. From the general KK metric presented in Eq.(\ref{5dfrt1}), $a=(e^{3\beta}e^{\gamma})^{1/4}$. Fig.\ref{Fig3} below shows the time evolution of $a$ for the present model, which reads

\begin{equation}\label{eqn:cp1}
a=e^{\frac{\lambda+\varpi(t,\lambda)}{\lambda^{2}t}}t^{1/2},
\end{equation}
with $\varpi(t,\lambda)\equiv \sqrt{\lambda[\lambda+2(8\pi+\lambda)t]}$. 

Note that, fundamentally, an expanding universe should have positive time derivatives for $a$, which is being respected in Fig.\ref{Fig3}. 

\begin{figure}[ht!]
\vspace{0.5cm}
\centering
\includegraphics[height=5.5cm,angle=00]{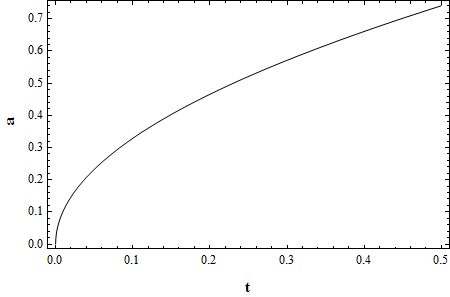}
\caption{Time evolution of the mean scale factor $a$ with $C_1=-1$ and $\lambda=-5\pi$.}
\label{Fig3}
\end{figure}

Note also that Eqs.(\ref{emtnc3})-(\ref{emtnc4}) have been applied for the dependence on $l$ in the calculation of $a$, as it was previously mentioned. The same approach will be applied in the calculation of the next cosmological parameters.

From the mean scale factor, one is able to obtain the Hubble parameter $H=\dot{a}/a$, which relates the recession velocity of galaxies with their distance through Hubble's law: $v=H(t)r$. From Eq.(\ref{eqn:cp1}), 

\begin{equation}\label{eqn:cp2}
H=\frac{2\varpi(t,\lambda)+\lambda[(\varpi(t,\lambda)+2)t+2]+16\pi t}{2\lambda\varpi(t,\lambda)t^{2}}.
\end{equation}
The evolution of $H$ through time is shown in Figure \ref{Fig4} below. One should note that the plot does not predict negative values for $H$, which would be a physical inconsistency in an expanding universe. Moreover, since $H\propto t_H^{-1}$, with $t_H$ being the Hubble time, $H$ must decrease with time, which can be observed in Fig.\ref{Fig4}.

\begin{figure}[ht!]
\vspace{0.5cm}
\centering
\includegraphics[height=5.5cm,angle=00]{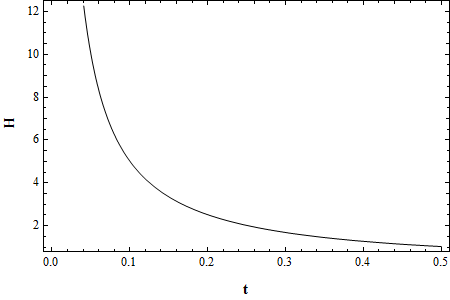}
\caption{Time evolution of the Hubble parameter $H$ with $C_1=-1$ and $\lambda=-5\pi$.}
\label{Fig4}
\end{figure} 

In cosmology is also common to work with the deceleration parameter $q=-\ddot{a}/(\dot{a}H)$, so that for negative values of $q$, the universe expansion is accelerating. Figure \ref{Fig5} below shows the behavior of $q$ through time from Eqs.(\ref{eqn:cp1})-(\ref{eqn:cp2}). Note that the negative values of $q$ for high values of $t$ are in accordance with recent cosmic acceleration predicted in \cite{riess/1998,perlmutter/1999,hinshaw/2013}.

\begin{figure}[ht!]
\vspace{0.5cm}
\centering
\includegraphics[height=5.5cm,angle=00]{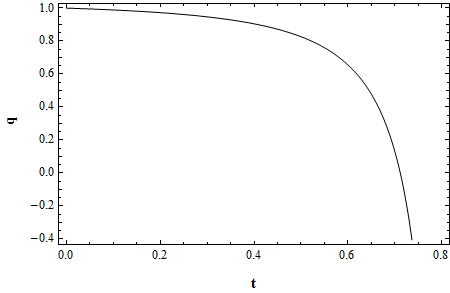}
\caption{Time evolution of the deceleration parameter $q$ with $C_1=-1$ and $\lambda=-5\pi$.}
\label{Fig5}
\end{figure} 

\section{Discussion}
\label{sec:dis}

In this work I have presented IMM solutions to 5D $f(R,T)$ theory from a general KK metric in which the coefficients depend on both time and extra coordinate. The $f(R,T)$ gravity model depends on a source term, representing the variation of the energy-momentum tensor with respect to the metric. Taking into account the covariant divergence of the energy-momentum tensor, I have obtained an equation for the evolution of the extra coordinate through time. Such an equation has revealed, in a natural form, the compactification of the fifth dimension (it should be noted that the compactification of the extra coordinate is usually imposed in KK models or even in 5D $f(R,T)$ models \cite{reddy/2013,sahoo/2014b,moraes/2014}, while here it has been obtained purely from the application of the non-conservation of the energy-momentum tensor, predicted by $f(R,T)$ theories). Furthermore, when the relation $l(t)$ is substituted in the cosmological parameters of the model, one obtains a projection of the fifth coordinate evolution in our observable 4D universe. Such a substitution is in accordance with recent observations of anisotropies in the cosmic microwave background temperature \cite{hinshaw/2013}, which indicates a negative deceleration parameter for the present universe dynamics. 

Another applications of IMM have been explored in \cite{halpern/2002,halpern/2001}. In these cases, instead of the $f(R,T)$ theory, the application was made to 5D anisotropic models. In \cite{halpern/2002}, solutions for the density and pressure of the universe in Bianchi type-II scenario presented exponential behavior. Note that solutions (\ref{5dfrt13})-(\ref{5dfrt14}) also present such a feature.

The behavior $\rho\propto t^{-1}$ in (\ref{5dfrt13}) is not a novelty in extradimensional models. In Randall-Sundrum braneworld model \cite{randall/1999}, which explains the hierarchy problem cited in Section \ref{sec:intro}, or even in its generalization \cite{sahni/2003}, the existence of a bulk in which our usual 4D universe (brane) is embedded yields an attenuation in the time evolution of $\rho$. Such an attenuation is predicted from the modified Friedmann Equations. While the standard 4D Friedmann Equation predicts a linear dependence on the matter-energy density $(H^{2}\propto\rho)$, in braneworld models the matter-energy density of the brane enters quadratically on the equation ($H^{2}\propto\rho^{2}$) (remind that the Hubble parameter is inversely proportional to the Hubble time). The present work corroborates, through Eq.(\ref{5dfrt13}), the prediction that extradimensional models should present an attenuation in the time evolution of $\rho$.  

Moreover, in solutions (\ref{5dfrt13})-(\ref{5dfrt14}), if one assumes $C_2=0$, note that the dependence on the extra coordinate vanishes. Such solutions describe a stiff matter dominated scenario, with EoS $p=\rho$. Some early universe models indicate that there may have existed a phase prior to that of radiation in which our universe dynamics was dominated by stiff matter (see, for instance, \cite{oliveira-neto/2011,rama/2009}, the latter being valid for any number of spatial dimensions). The presence of stiff matter in cosmological models may explain the baryon asymmetry and the density perturbations of the right amplitude for the large scale structure formation in the universe \cite{zeldovich/1972,joyce/1998}. It may also play an important role in the spectrum of gravitational waves generated in the inflation era \cite{sahni/2001}. These points shall be carefully investigated and reported in future works. 

In Subsection \ref{ss:to}, I have derived IMM time (only)-dependent solutions from (\ref{5dfrt7})-(\ref{5dfrt9}). The same sort of approach was applied in \cite{wesson/1992b,halpern/2002}. The relation $\rho\propto t^{-2}$ of $\Lambda$CDM cosmological model is retrieved in solution (\ref{time2}). Such a retrieval was also obtained from IMM application in \cite{halpern/2001}.

In Section \ref{sec:emtnc} I have derived a relation between $l$ and $t$ from the non-conservation of the energy-momentum tensor, which is a fundamental characteristic of $f(R,T)$ theories. Such a relation reveals the compactification of the extra coordinate. A difficulty with compactification is that it cannot be imposed indiscriminately on whichever dimensions one likes. This model predicts compactification (check Fig.\ref{Fig1}) instead of imposing it. 

Besides, the functional form of $l(t)$ has revealed a very substantial feature about the 4D universe dynamics. The substitution of Eqs.(\ref{5dfrt10}),(\ref{5dfrt12}),(\ref{emtnc3}) and (\ref{emtnc4})\footnote{For the values chosen for the constants of the model, it seemed reasonable to work with Eq.(\ref{emtnc4}) instead of Eq.(\ref{emtcn5}). The latter might be explored in future works by assuming different values for $\lambda$, $C_1$ and $C_2$.} in $a=(e^{3\beta}e^{\gamma})^{1/4}$ reduces the cosmological parameters of the universe to the usual 4D space-time. Such a substitution, for which, for the sake of simplicity I have used the same values for $C_1$ and $\lambda$ used in Fig.\ref{Fig1}, has disclosed the recent accelerated expansion of the observable universe, through a negative deceleration parameter for high values of $t$. Such a dynamical behavior might be interpreted as a geometrical effect obtained from the shrinking of the extra dimension. Note that in order to obtain an accelerated expansion, the present model did not assume the existence of any kind of exotic fluid or scalar field. Instead, the relation which exposes a natural shrinking of the extra dimension also display naturally an accelerated cosmic expansion in 4D when substituted in the cosmological parameters.

Therefore, the intriguing acceleration of the universe expansion could be justified purely in a geometrical perspective, since the compactification of the extra dimension yields $q<0$, which accounts for the present scenario of the universe dynamics (check \cite{hinshaw/2013}). In fact, the unification of $f(R,T)$ theory with KK cosmology through IMM application highlights the importance that the geometry of a 5D universe has in our 4D observable universe.   

One might wonder the reason for the cosmological solutions of the present model do not recover standard gravity solutions when $\lambda=0$, as originally proposed in \cite{harko/2011}. One should note that such a recovering is predicted in the 4D theory. The presence of extradimensional terms makes impossible to retrieve standard gravity this way.\\

{\bf Acknowledgements} I would like to thank CAPES for financial support. I would also like to thank Dr. E.S. Pereira for some valuable comments on the cosmological interpretation of Eqs.(\ref{emtnc3})-(\ref{emtcn5}).


\end{document}